\documentstyle[12pt]{article}
\newcommand{\be}{\begin{equation}}
\newcommand{\ee}{\end{equation}}

\date{September 1998}
\begin{document}
\begin{center}
{\large {\bf On realizations of ``nonlinear'' Lie algebras by\\

differential operators}}\\
\vspace{0.5in}
J. BECKERS, $^{a)}$\footnote{E-mail : Jules.Beckers@ulg.ac.be} Y.
BRIHAYE, $^{b)}$\footnote{E-mail : Yves.Brihaye@umh.ac.be} and N.
DEBERGH
$^{a)}$\footnote{Chercheur, Institut Interuniversitaire des Sciences
Nucl\'eaires, Bruxelles}

\vspace{0.3in}
a) {\it Theoretical and Mathematical Physics,\\
Institute of Physics (B5)\\
University of Li\`ege,\\
B-4000 Li\`ege 1 (Belgium)}\\
\vspace{0.2in}
b) {\it Department of Mathematical Physics,\\
University of Mons-Hainaut,\\
Av. Maistriau,\\
B-7000 Mons (Belgium)}
\end{center}
\vspace{0.5in}
\begin{abstract}
We study realizations of {\it polynomial} deformations of
the $\rm sl(2,\Re \rm)$-Lie algebra in terms of differential operators
strongly related to bosonic operators. We also distinguish their
finite- and infinite-dimensional representations. The linear, quadratic
and cubic cases are explicitly visited but the method works for
arbitrary degrees in the polynomial functions. Multi-boson
Hamiltonians are studied in the context of these ``{\it nonlinear}'' Lie
algebras and some examples dealing with {\it quantum optics} are pointed
out.
\end{abstract}
\newpage
\section{Introduction}
\hspace{0.5cm}	Lie groups and algebras [1] have a very large
importance in quantum physics [2] in particular : they are associated
with symmetry properties of physical systems and always improve the
understanding of a lot of applications. In fact, such developments are
mainly based on {\it linear} Lie algebras, but it becomes more and more
evident that there is no physical reason for symmetries to be only
linear ones, leading to the observation that Lie theory therefore may
be too restrictive. Very recent studies have visited important classes
of {\it nonlinear} symmetries. Among these, let us only mention the
nonlinear finite W-symmetries [3] described by {\it nonlinear Lie
algebras} that we define as ``gene\-ralizations of ordinary Lie algebras
containing different order pro\-ducts of the generators on the
right-hand side of the defining brackets without violating Jacobi
identities'' [4]. Another recent approach [5, 6] has consi\-dered
nonlinear angular momentum theories fundamentally developed from
nonlinear extensions of the (real form $\rm sl(2,\Re \rm)$ of the)
complex Lie-Cartan algebra
$\rm A_1$. These two contents [3, 5, 6] are not independent [7] and
cover an important set of contributions on the corresponding {\it
representations} of nonlinear Lie algebras [4-14] where, more
particularly, {\it quadratic} and {\it cubic} nonlinearities have been
extensively handled.

The main purpose of this study is to discuss and put forward possible
realizations of nonlinear algebras by {\it differential} operators,
such representations being also of primary interest in connection with
physical applications described by typical quantum Hamiltonians. As an
example, we choose to study {\it polynomial} deformations of the
linear $\rm sl(2,\Re \rm)$-algebra written in the form
\be
\label{1}
\rm [J_0, J_{\pm}] = \pm J_{\pm} ,
\ee
\be
\label{2}
\rm [J_+,J_-] = P(J_0) ,
\ee
where, as usual, $\rm J_\pm$ are the well known raising (+) and
lowering (-) operators (physically important in the angular momentum
theory [15]) and $\rm J_0$ the diagonal one, $\rm P(J_0)$ being a
polynomial function of $\rm J_0$ with a generally finite degree
$\Delta$. Particular nonlinear $\rm sl(2,\Re \rm)$-algebras have
already characterized physical applications in quantum mechanics [3]
but also in Yang-Mills type gauge theories, inverse scattering,
conformal field theories [16, 17], quantum optics [18, 19], etc.

The interest of realizations in terms of linear differential
operators is evidently stressed by the fact that position and
momentum operators are strongly related to linear combinations of
(harmonic) oscillator creation $\rm (a^\dagger)$ and annihilation (a)
operators so that the so-called {\it bosonic} realizations of
nonlinear Lie algebras appear as very rich information. Let us here
only recall the Heisenberg algebra characterized by the nonzero
commutation relation (in a 1-dimensional space)
\begin{eqnarray}
\label{eq3}
\rm \left[{{d \over dx}\ ,\ x}\right]=1
\end{eqnarray}
equivalent to the oscillator algebra
\begin{eqnarray}
\label{eq4}
\rm \left[a,a^\dagger \right]=1
\end{eqnarray}
with the corresponding number operator given here by the dilation
operator
\begin{eqnarray}
\label{eq5}
\rm D\equiv \rm x\ {d \over dx} \ \ .
\end{eqnarray}
Such bosonic representations have already been studied for a long time
in connection with (linear) Lie algebras [20, 21] but we want to
tackle nonlinear ones as already mentioned.

The contents are then distributed as follows. In Section 2, we propose
a differential realization of the three generators $\rm \left({{J}_{\pm
}\ ,\ {J}_{0}}\right)$ satisfying the typical commutation relations
(1.1) and (1.2) and we discuss three specific polynomial contexts :
the (trivial) linear case (\S 2.A), the quadratic context (\S 2.B) which
is a first nontrivial (nonlinear) case with several physical
motivations [4, 8, 10-13] and the cubic case (\S 2.C), another specific
finite W-algebra once again very well visited [5-7, 9, 13, 22, 23] in
connection with diffe\-rent physical applications. An important
discussion will also take place in that sec\-tion~: we want to
distinguish the possible {\it finite-dimensional} representations of
the different algebras. Moreover, we will only be interested in
realizations which could be handled in the future, formal ones being
then discarded. In Section 3, as an application, we visit families of
multi-boson Hamiltonians [19] and point out some of them admitting
typically such nonlinear Lie algebras as spectrum generating algebras.

Our units are taken with the
constant $\rm \hbar$ equal to unity.
\section{Differential realizations of the generators}
\setcounter{equation}{0}
\hspace{0.5cm}	Let us realize the nonlinear algebra (1.1)-(1.2) in
terms of differential operators depending on a real variable x. Such
commutation relations suggest that these operators have the following
forms :
\begin{eqnarray}
\rm {J}_{+}={x}^{N}\ F(D)\ \ \ ,\ \ \ {J}_{-}=G(D)\ {{d}^{N} \over
d{x}^{N}}\ \ \ ,\ \ \ {J}_{0}={1 \over N}\ (D+c) ,
\label{eq2.1}
\end{eqnarray}
where c is a constant, D is given by eq. (1.5) and N=1,2,3...\ .
Besides the starting property (1.3), let us mention the following
identities and notations which are very useful in order to manipulate
the above operators :
\begin{eqnarray}
\rm \left[{D,{x}^{N}}\right]=N\ {x}^{N}\ \ \ ,\ \ \ \left[{{{d}^{N}
\over d{x}^{N}},D}\right]=N\ {{d}^{N} \over d{x}^{N}}
\label{eq2.2}
\end{eqnarray}
and
\begin{eqnarray}
\rm {{d}^{N} \over d{x}^{N}}\ {x}^{N}=\prod\limits_{\rm j=1}^{\rm N} \rm
(D+j)={(D+N)! \over D!} ,
\label{eq2.3}
\end{eqnarray}
\begin{eqnarray}
\rm {x}^{N}\ {{d}^{N} \over d{x}^{N}}=\prod\limits_{\rm j=0}^{\rm N-1}
\rm (D-j)={D! \over (D-N)!} .
\label{eq2.4}
\end{eqnarray}
If the relations (1.1) are easily satisfied, the one given in eq.
(1.2) asks for a general constraint written as
\begin{eqnarray}
\rm F(D-N)G(D-N){D! \over (D-N)!}-F(D)G(D) {(D+N)! \over
D!}=P\left({{1 \over N}(D+c)}\right) .
\label{eq2.5}
\end{eqnarray}
If $\rm P(J_0)$ in eq. (1.2) is a polynomial of degree $\Delta$ in $\rm
J_0$, we evidently should have $\Delta = \rm deg_D(FG) + N - 1$.

Here we want to point out that the following transformation
\begin{eqnarray}
\rm {\widehat{J}}_{+}={J}_{+}U\ \ \ ,\ \ \
{\widehat{J}}_{-}={U}^{-1}{J}_{-}\ \ \ ,\ \ \
{\widehat{J}}_{0}={J}_{0}
\label{eq2.6}
\end{eqnarray}
defines an automorphism of the realization (2.1) if U is a function of
D only. Such an automorphism allows us to simplify the functions F(D)
and (or) G(D). In the following, we exploit this freedom without loss
of generality. It is thus understood that all the realizations
explicitly presented here will be valid up to an automorphism of the
form (2.6). So, let us fix our choice on
\begin{eqnarray}
\rm G(D)=1
\label{eq2.7}
\end{eqnarray}
in the constraint (2.5) and let us apply it to specific cases which are
particularly meaningful in connection with physical applications [3,
4, 16, 17] : we want to treat successively the three cases of the ({\it
linear}) algebra $\rm sl(2,\Re \rm)$ and the ({\it nonlinear})
quadratic and cubic deformations of this structure.
\subsection{The linear algebra $\rm {\bf sl(2,{\bf \Re} \rm)}$}

\hspace{0.5cm} This case corresponds to the well known example where
\begin{eqnarray}
\rm P({J}_{0})=2{J}_{0}={2 \over N}\ (D+c)
\label{eq2.8}
\end{eqnarray}
subtending all the very important consequences of the angular momentum
theory [15] and its finite dimensional representations but dealing
with a specific basis other than the one in which we are interested
here. Let us also mention a recent study [21] on this case which can
be strongly related to ours up to specific periodic conditions imposed
in that work [21]. Here the function F(D) is chosen as
\begin{eqnarray}
\rm F(D)=-{D! \over {N}^{2}(D+N)!}\ (D+{\lambda }_{1})(D+{\lambda
}_{2})
\label{eq2.9}
\end{eqnarray}
where ${\lambda }_{1}$ and ${\lambda }_{2}$ are x-independent
parameters. The constraint (2.5) with the functions (2.7) and (2.9)
leads to
\begin{eqnarray}
\rm {\lambda }_{1}+{\lambda }_{2}=2c+N
\nonumber
\end{eqnarray}
so that in terms of a single parameter $\lambda$ [21] we get
\begin{eqnarray}
\rm {\lambda }_{1}=c+{N \over 2}+\lambda \rm \ \ and\ \ {\lambda
}_{2}=c+{N \over 2}-\lambda \ \  .
\label{eq2.10}
\end{eqnarray}
Then the discussion of the function F(D) $\equiv$ (2.9) starts with
different N-values but by limiting ourselves to nonsingular
realizations of the three gene\-rators (2.1).

Here only the values N=1,2 are admissible ones in that sense and we
immediately get that, {\it for N=1},
\begin{eqnarray}
\rm \lambda \rm =\pm ({1 \over 2}-c)\ \Rightarrow \rm \ F(D)=-(D+2c)
\label{eq2.11}
\end{eqnarray}
and, {\it for N=2},
\begin{eqnarray}
\rm \lambda \rm =\pm {1 \over 2}\ ,\ c={1 \over 2}\ \Rightarrow \rm \
F(D)=-{1 \over 4} \ \ .
\label{eq2.12}
\end{eqnarray}
Let us give once these explicit realizations in the two cases : we
respectively have
\begin{eqnarray}
\rm {J}_{+}=-x\left({x{d \over dx}+2c}\right)\ \ \ ,\ \ \ {J}_{-}={d
\over dx}\ \ \ ,\ \ \ {J}_{0}=x{d \over dx}+c
\label{eq2.13}
\end{eqnarray}
and
\begin{eqnarray}
\rm {J}_{+}=-{{x}^{2} \over 4}\ \ \ ,\ \ \
{J}_{-}={{d}^{2} \over {dx}^{2}}\ \ \ ,\ \ \ {J}_{0}={1 \over
2}\left({x{d \over dx}+{1 \over 2}}\right)  .
\label{eq2.14}
\end{eqnarray}

Let us also point out that, if we ask for finite-dimensional
representations, we have to consider P(n) (the (n+1)-dimensional
vector space of polynomials of degree at most n in the x-variable) and
to preserve it under the action of the generator $\rm {J}_{+}$
in particular. This requires in particular that
\begin{eqnarray}
\rm {J}_{+}{x}^{n}=F(n)\ {x}^{n+N}=0
\label{eq2.15}
\end{eqnarray}
and thus that
\begin{eqnarray}
\rm F(n)=F(n-1)=...=F(n-N+1)=0 .
\label{eq2.16}
\end{eqnarray}
{\it Only} the case N=1
with  $\rm \displaystyle c=-{n \over 2}$ is permitted, the
corresponding generators (2.13) being already extensively used in the
construction of quasi exactly solvable equations [24, 25].

\subsection{The quadratic algebra}

\hspace{0.5cm} If we require (with arbitrary $\alpha_i$'s, $i=1, 2, 3$
up to $\alpha_3 \neq 0$)
\begin{eqnarray}
\rm P({J}_{0})= \alpha_1 + \alpha_2{J}_{0}+ \alpha_3 {\rm J}_{\rm
0}^{\rm 2}
\
\ ,
\label{eq2.17}
\end{eqnarray}
we are concerned with an already well visited nonlinear algebra [8,13]
with specific physical interests [4] if $\alpha_1 = 0, \alpha_2 = 2,
\alpha_3 = 4 \alpha$ {\it or} with a $W_3^{(2)}$-algebra [3] if
$\alpha_2 = 0$ and $\alpha_3 > 0$. Let us apply our considerations and
search for functions
\begin{eqnarray}
\rm F(D)=-d\ {D! \over (D+N)!}\ (D+{\lambda }_{1})(D+{\lambda
}_{2})(D+{\lambda }_{3})
\label{eq2.18}
\end{eqnarray}
satisfying the condition (2.5) with the entries (2.7) and (2.17). This
leads to the information
\begin{eqnarray}
\rm d={\alpha_3  \over \rm 3{N}^{3}}\ \ \ , \ \ \ {\lambda
}_{1}+{\lambda }_{2}+{\lambda }_{3}={3N\alpha_2 \over 2\alpha_3 }+3c+{3N
\over 2} , \nonumber \\
\rm {\lambda }_{1}{\lambda }_{2}+{\lambda }_{1}{\lambda }_{3}+{\lambda
}_{2}{\lambda }_{3}={{N}^{2} \over 2}+3cN+{{3N}^{2}\alpha_2 \over
2\alpha_3 }+{3Nc\alpha_2 \over \alpha_3 }+3{c}^{2} + 3N^2
\frac{\alpha_1}{\alpha_3} \ .
\label{eq2.19}
\end{eqnarray}
Due to the specific form of our functions (2.18), we have only three
admissible values N=1,2,3. We get for {\it N=1}
\begin{eqnarray}
&&d = \frac{\alpha_3}{3} \; , \; \lambda_1 = 1 \; , \; \lambda_2 =
\frac{3}{2} \rm c + \frac{1}{4} + \frac{3}{4} \frac{\alpha_2}{\alpha_3}
+\frac{\epsilon}{2}, \nonumber \\
&&\lambda_3 =
\frac{3}{2} \rm c + \frac{1}{4} + \frac{3}{4} \frac{\alpha_2}{\alpha_3}
-\frac{\epsilon}{2},\nonumber \\
&& \epsilon = (-3\rm c^2 + 3\rm c + \frac{1}{4} +
\frac{9}{4}
\frac{\alpha_2^2}{\alpha_3^2} - \frac{\alpha_2}{\alpha_3} (\frac{3}{2}
- 3\rm c) - 12 \frac{\alpha_1}{\alpha_3})^{\frac{1}{2}}
\ \
\label{eq2.20}
\end{eqnarray}
so that the corresponding differential realization is
\begin{equation}
\rm J_0 = \rm D + \rm c\; , \; \rm J_- = \frac{\rm d}{\rm {dx}}\; , \;
\rm J_+ = -
\frac{\alpha_3}{3} \rm x (\rm D + \lambda_2) (\rm D + \lambda_3).
\label{eq2.21}
\end{equation}
It will correspond to finite-dimensional representations of
dimension $(n+1)$ iff $\lambda_2 = -n$ or $\lambda_3 = -n$.
\par
{\it For N=2}, we have
\begin{eqnarray}
\rm d = \frac{\alpha_3}{24}\; , \; {\lambda }_{1}=1\ \ \ ,\ \ \ {\lambda
}_{2}=2\
\
\ ,\
\
\ {\lambda }_{3}=3 (\rm c + \frac{\alpha_2}{\alpha_3})
\label{eq2.22}
\end{eqnarray}
and the realization writes
\begin{equation}
\rm J_0 = \frac{1}{2} (\rm D + \rm c)\; , \; \rm J_- =
\frac{\rm d^2}{\rm {dx}^2}\; ,
\; \rm J_+ = -
\frac{\alpha_3}{24} \rm x^2 (\rm D + 3\rm c + 3
\frac{\alpha_2}{\alpha_3}).
\label{eq2.23}
\end{equation}
leading to infinite-dimensional representations only.
\par
Finally, {\it for N=3}, we have
\begin{eqnarray}
&&\rm d = \frac{\alpha_3}{81} \; , \; \lambda_1 = 1\; , \; \lambda_2 = 2
\; , \; \lambda_3 = 3, \nonumber \\
&&\frac{\alpha_2}{\alpha_3} = - \frac{2}{3}\rm c + \frac{1}{3} \; , \;
\frac{\alpha_1}{\alpha_3} = \frac{1}{9} \rm c^2 - \frac{1}{9} \rm c +
\frac{2}{27}
\label{eq2.24}
\end{eqnarray}
giving, once again, infinite-dimensional representations through the
differential realization
\begin{equation}
\rm J_0 = \frac{1}{3} (\rm D + \rm c)\; , \; \rm J_- =
\frac{\rm d^3}{\rm {dx}^3}\; ,
\; \rm J_+ = -
\frac{\alpha_3}{81}\rm x^3.
\label{eq2.25}
\end{equation}
\par
Let us end this subsection by mentioning that the realizations and
representations of the {\it quadratic} algebra [4] characterized by
\begin{equation}
\rm P(\rm J_0) = 2 \rm J_0 + 4 \alpha \rm J_0^2
\label{eq2.26}
\end{equation}
are readily obtained by fixing the $\alpha_i$'s as already specified.

\subsection{The cubic or Higgs algebra}

\hspace{0.5cm} The Higgs algebra [22] has also been intensively
exploited till now, in connection {\it either} with specific models
[13,18,22,23] {\it or} through technical relations with W-algebras [3,7]
as well as with the study of its irreducible representations [5,6,9].
Here it is characterized by the polynomial function
\begin{eqnarray}
\rm P({J}_{0})=2{J}_{0}+8\beta {\rm J}_{\rm 0}^{\rm 3}\rm \ \ ,
\label{eq2.27}
\end{eqnarray}
$\beta (\neq 0)$ being very often interpreted as a deformation
parameter. The functions F(D) corresponding to the previous expressions
(2.9) and (2.18) take the form
\begin{eqnarray}
\rm F(D)=-f\ {D! \over (D+N)!}(D+{\lambda }_{1})(D+{\lambda
}_{2})(D+{\lambda }_{3})(D+{\lambda }_{4})\ \ .
\label{eq2.28}
\end{eqnarray}
We get the information :
\begin{eqnarray}
\rm f={2\beta  \over {\rm N}^{\rm 4}}\ \ ,\ \ {\lambda }_{1}+{\lambda
}_{2}+{\lambda }_{3}+{\lambda }_{4}&=& \rm 4c+2N\ \ , \nonumber \\
\rm {\lambda }_{1}{\lambda }_{2}+{\lambda }_{1}{\lambda }_{3}+{\lambda
}_{2}{\lambda }_{3}+{\lambda }_{1}{\lambda }_{4}+{\lambda }_{2}{\lambda
}_{4}+{\lambda }_{3}{\lambda }_{4}&=& \rm {N}^{2}+6cN+6{c}^{2}+{{N}^{2}
\over 2\beta }\ \ , \nonumber \\
{\lambda }_{1}{\lambda }_{2}{\lambda }_{3}+{\lambda }_{1}{\lambda
}_{2}{\lambda }_{4}+{\lambda }_{1}{\lambda }_{3}{\lambda }_{4}+{\lambda
}_{2}{\lambda }_{3}{\lambda
}_{4}\nonumber \\
= \rm 2c{N}^{2}+6{c}^{2}N+4{c}^{3}&+& \rm {c{N}^{2} \over \beta }+{1
\over 2\beta }{N}^{3}
\label{eq2.29}
\end{eqnarray}
and have to discuss the cases N=1,2,3 and 4. Here are the results {\it
for N=1} :
\begin{eqnarray}
\rm {\lambda }_{1}=1\ \ ,\ \ {\lambda }_{2}={1 \over 2}+c+{1
\over 2}\sqrt {1+4c-4{c}^{2}-{2 \over \beta }}\ \ , \nonumber \\
\rm {\lambda }_{3}={1 \over 2}+c-{1 \over 2}\sqrt
{1+4c-4{c}^{2}-{2
\over
\beta }}\ \ ,\ \ {\lambda }_{4}=2c\ \ ;
\label{eq2.30}
\end{eqnarray}
{\it N=2} :
\begin{eqnarray}
{\lambda }_{1}=1\ \ ,\ \ {\lambda }_{2}=2\ \ ,
\label{eq2.31}
\end{eqnarray}
\begin{eqnarray}
\mit if\rm \ \ c={1 \over 2}\ ,\ {\lambda }_{3}={3 \over 2}+{1 \over
2}\sqrt {7-{8 \over \beta }}\ ,\ {\lambda }_{4}={3 \over 2}-{1 \over
2}\sqrt {7-{8 \over \beta }}\ ,
\label{eq2.32}
\end{eqnarray}
\begin{eqnarray}
\mit or\rm \ \ c={1 \over 2}+{1 \over 2}\sqrt {3-{4 \over \beta }}\ ,\
{\lambda }_{3}=1+\sqrt {3-{4 \over \beta }}\ ,\ {\lambda }_{4}=2+\sqrt
{3-{4 \over \beta }}\ ,
\label{eq2.33}
\end{eqnarray}
\begin{eqnarray}
\mit or\rm \ \ c={1 \over 2}-{1 \over 2}\sqrt {3-{4 \over \beta }}\ ,\
{\lambda }_{3}=1-\sqrt {3-{4 \over \beta }}\ ,\ {\lambda }_{4}=2-\sqrt
{3-{4 \over \beta }}\ ;
\label{eq2.34}
\end{eqnarray}
{\it N=3} :
\begin{eqnarray}
\rm {\lambda }_{1}=1\ ,\ {\lambda }_{2}=2\ ,\ {\lambda }_{3}=3\ ,\
{\lambda }_{4}=4c\ ,
\label{eq2.35}
\end{eqnarray}
\begin{eqnarray}
\mit if\rm \ \ c=0\Rightarrow{\lambda }_{4}=0\ ,\ \beta \rm ={9
\over 4}\ ;
\label{eq2.36}
\end{eqnarray}
\begin{eqnarray}
\mit or\rm \ \ c={1 \over 2}\Rightarrow {\lambda }_{\rm 4}\rm =2\ ,\
\beta \rm ={9 \over 7}\ ;
\label{eq2.37}
\end{eqnarray}
\begin{eqnarray}
\mit or\rm \ \ c=1\Rightarrow {\lambda }_{\rm 4}\rm =4\ ,\ \beta \rm
={9 \over 4}\ ;
\label{eq2.38}
\end{eqnarray}
{\it N=4} :
\begin{eqnarray}
\rm {\lambda }_{1}=1\ ,\ {\lambda }_{2}=2\ &,&\ {\lambda }_{3}=3\ ,\
{\lambda }_{4}=4\ : \nonumber \\
\rm c={1 \over 2}\ &,&\ \beta \rm ={16 \over 11}\ .
\label{eq2.39}
\end{eqnarray}
Finite-dimensional representations are only possible for N=1 and 2. If
N=1, ${\lambda }_{1}=1$ and ${\lambda }_{2}$= -n are fixed but ${\lambda
}_{3}$ and ${\lambda}_{4}$ can take three different sets of values in
correspondence with well definite values of c as follows :
\begin{eqnarray}
\mit if\ \rm c=-{n \over 2}\ ,\ {\lambda }_{3}={1 \over 2}-{n \over
2}+{{R}_{1} \over 2}\ ,\ {\lambda }_{4}={1 \over 2}-{n \over
2}-{{R}_{1} \over 2}
\label{eq2.40}
\end{eqnarray}
with
\begin{eqnarray}
\rm {R}_{1}={\left({1-2n-{n}^{2}-{2 \over \beta }}\right)}^{1/2}\ ;
\label{eq2.41}
\end{eqnarray}
\begin{eqnarray}
\mit if\ \rm c=-{n \over 2}+{1 \over 2}{R}_{2}\ ,\ {\lambda
}_{3}=-n+{R}_{2}\ ,\ {\lambda }_{4}=1+{R}_{2}
\label{eq2.42}
\end{eqnarray}
with
\begin{eqnarray}
\rm {R}_{2}={\left({-{n}^{2}-2n-{1 \over \beta }}\right)}^{1/2}\ ;
\label{eq2.43}
\end{eqnarray}
\begin{eqnarray}
\mit if\ \rm c=-{n \over 2}-{1 \over 2}{R}_{2}\ ,\ {\lambda
}_{3}=-n-{R}_{2}\ ,\ {\lambda }_{4}=1-{R}_{2}\ .
\label{eq2.44}
\end{eqnarray}
Let us point out that these three families correspond to the three
ones obtained in the angular momentum basis [5] associated with n=2j.

If N=2, ${\lambda}_{1}=1$, ${\lambda}_{2}$=2 and ${\lambda}_{3}$=-n,
${\lambda}_{4}$=-n+1 are determined so that we get fixed values for c
and $\beta$ given by
\begin{eqnarray}
\rm c=-{n \over 2}\ \ ,\ \ \beta \rm =-{4 \over {n}^{2}+2n-2}
\label{eq2.45}
\end{eqnarray}
leading to supplementary representations recently obtained in [6].

The method developed here can evidently be applied to any
degree of nonlinearities in the polynomials characterizing eq. (1.2)
depending on specific questions and applications we want to consider.
In conclusion of this section, let us consider the q-deformation
${\cal U}_{\rm q}\left({\rm sl(2,\Re)}\right)$ of $\rm sl(2,\Re)$ which
corresponds to the commutation relation (1.2) written on the form
\begin{eqnarray}
\rm \left[{{J}_{+},{J}_{-}}\right]=\left[{2{J}_{0}}\right]
\label{eq2.46}
\end{eqnarray}
where the bracket refers to the usual notation [26]
\begin{eqnarray}
\rm \left[{x}\right]={{q}^{x}-{q}^{-x} \over q-{q}^{-1}}\ \ \ ,\ \ \
q={e}^{\gamma }\ \ ,
\label{eq2.47}
\end{eqnarray}
leading to
\begin{eqnarray}
\rm \left[{2{J}_{0}}\right]=\left[{{2 \over N}(D+c)}\right]={1 \over sh\
\gamma }\ sh\ {2\gamma  \over \rm N}(D+c)
\label{eq2.48}
\end{eqnarray}
where $\gamma$ characterizes the deformation parameter of this {\it
quantum} algebra. In our differential realization (2.1) with the
condition (2.7), the new functions F(D) take the form
\begin{eqnarray}
\rm F(D)=-{1 \over 2\ s{h}^{2}\ \gamma }\ {D! \over (D+N)!}\
ch\left({{2\gamma  \over \rm N}(D+c)+\gamma }\right)
\label{eq2.49}
\end{eqnarray}
also rewritten [27] as
\begin{eqnarray}
\rm F(D)=-{1 \over 2\ s{h}^{2}\ \gamma }\ {D! \over (D+N)!}\
\prod\limits_{\rm k=1}^{\infty } \left({\rm 1+{\left({{2\gamma \rm
(2D+2c+N \over N(2k+1)\pi }}\right)}^{2}}\right)\rm \ .
\label{eq2.50}
\end{eqnarray}
No admissible cases can thus be discussed and all the
representations are infinite-dimensional (as expected), c being still
a free parameter.

\section{Some examples from quantum optics}
\setcounter{equation}{0}
\hspace{0.5cm}	A very recent result [18] concerning a location of
fundamental super\-sym\-metry [28] in multi-boson Hamiltonians [19]
suggests to learn if our differential approach can be useful through
the exploitation of nonlinear Lie algebras seen as spectrum generating
algebras [29] of these Hamiltonians as it was the case in quantum optics
in particular.
\par
We thus propose to consider the following family of Karassiov-Klimov
Hamiltonians [19] given by
\begin{equation}
H = \omega_1 \rm a_1^{\dagger} a_1 + \omega_2 \rm a_2^{\dagger} a_2 +
\rm g (a_1^{\dagger})^s a_2^r + \bar {\rm g} a_1^s (a_2^{\dagger})^r
\end{equation}
where $0 \leq r \leq s$, $\omega_1$ and $\omega_2$ referring to
angular frequencies of two {\it independent} harmonic oscillators
characterized by annihilation ($a_1, a_2$) and creation
($a_1^{\dagger}, a_2^{\dagger}$) operators verifying the usual
Heisenberg relations (1.4).
\par
With the definitions
\begin{equation}
\rm R_0 \equiv \frac{1}{r+s} (ra_1^{\dagger} a_1 + sa_2^{\dagger} a_2) =
\frac{1}{r+s} (r N_1 + s N_2),
\end{equation}
\begin{equation}
\rm J_0 \equiv \frac{1}{r+s} (a_1^{\dagger} a_1-a_2^{\dagger} a_2)\; ,
\; \rm J_+ \equiv (a_1^{\dagger})^s a_2^r \; , \; \rm J_- = a_1^s
(a_2^{\dagger})^r,
\end{equation}
the Hamiltonian (3.1) becomes
\begin{equation}
H = (\omega_1 + \omega_2) \rm R_0 + (\omega_1 s - \omega_2 r) \rm J_0 +
g \rm J_+ + \bar g \rm J_-
\end{equation}
and we already notice that
\begin{equation}
[\rm R_0 , \rm J_0] = [\rm R_0 , \rm J_{\pm}] = 0 \; , \; [\rm J_0 ,
\rm J_{\pm}] =
\pm \rm J_{\pm}
\end{equation}
for arbitrary values of $r$ and $s$. The Hamiltonian (3.1) suggests an
infinite-dimensional basis $\{ \mid n_1, n_2 \rangle \; \;  n_1, n_2 =
0, 1, 2, ...\}$ for which the eigenvalues of $\rm R_0$ (noted by $j$ in
the following) are of the form $j$ = ($r n_1 + s n_2$)/($r+s$).
\subsection{Towards the (cubic) Higgs algebra}
\hspace{5mm}
In connection with the results [18] enhancing the Higgs algebra as a
spectrum generating algebra of the Hamiltonian (3.4), we have to add
to eqs.(3.5) the requirement that the commutator $[\rm J_+ , \rm J_-]$
is up to a renormalization of $\rm J_{\pm}$, of the form (1.2) with
(2.27). It was realized in [18] that this is possible only for
$r=s=2$. Here the commutator reads
\begin{equation}
[\rm J_+ , \rm J_-] = - 64 \rm J_0^3 + 8 \rm J_0 (2 \rm R_0^2 + 2 \rm
R_0 - 1) .
\end{equation}
Comparison of (2.27) with (3.6) further indicates that the parameter
$\beta$ should be of the form
\begin{equation}
\beta = -\frac{4}{4j^2 + 4j - 2} \; \; , \; j = 0, \frac{1}{2}, 1, ...
,
\end{equation}
which, in passing we note, is very similar to the condition (2.45)
obtained in our classification of finite-dimensional representations
of the Higgs algebra.
\par
After calculations, one can show that, out of the four cases $\rm N= 1,
2, 3, 4$, only the $\rm N =1$- and $\rm N=2$- contexts (see
(2.30)-(2.34)) are possible.
\par
We first discuss the three solutions associated with the context {\it
when $N = 2$}:
\begin{equation}
 if \; \; \; \; \; \;  c = \frac{1}{2} \; , \; \lambda_3 = \frac{3}{2} +
\frac{1}{2} (8j^2+8j+3)^{\frac{1}{2}} \; , \; \lambda_4 = 3 -
\lambda_3,
\end{equation}
we have
\begin{equation}
\rm J_0 = \frac{1}{2}(\rm D + \frac{1}{2}) \; , \; \rm J_- = \frac{\rm
d^2}{\rm {dx}^2} \; , \; \rm J_+ = \rm x^2 (\rm D + \lambda_3)(\rm D +
3 - \lambda_3);
\end{equation}
\begin{equation}
 if \; \; \; \; \; \;  c = j+1 \; , \; \lambda_3 = 2j+2 \; , \;
\lambda_4 = 2j+3,
\end{equation}
we obtain
\begin{equation}
\rm J_0 = \frac{1}{2}(\rm D + j + 1) \; , \; \rm J_- = \frac{\rm
d^2}{\rm {dx}^2} \; , \; \rm J_+ = \rm x^2 (\rm D + 2j + 2)(\rm D +
2j + 3);
\end{equation}
\begin{equation}
 if \; \; \; \; \; \;  c = -j \; , \; \lambda_3 = -2j \; , \; \lambda_4
= -2j+1,
\end{equation}
we get
\begin{equation}
\rm J_0 = \frac{1}{2}(\rm D - j) \; , \; \rm J_- = \frac{\rm
d^2}{\rm {dx}^2} \; , \; \rm J_+ = \rm x^2 (\rm D - 2j)(\rm D -
2j + 1).
\end{equation}
Only the realization (3.13) -- which exactly corresponds to the
discussion leading to eqs. (2.45) -- is of finite dimension while
those given by eqs. (3.9) and (3.11) are infinite-dimensional ones.
\par
Let us now insert such realizations inside the Hamiltonians (3.4) with
the {\it ideal} conditions $\omega_1 = \omega_2 = \omega$, $r=s=2$ and
$g = \bar g$ in order to determine if the above three contexts are or
are not typical ones of the {\it linear} algebra $sl(2,\Re)$ or of
powers of its generators already obtained in eqs. (2.13) and (2.14).
\par
With the finite-dimensional realization (3.13), we readily find the
corresponding Hamiltonian on the form
\begin{equation}
H = 2 \omega j + g (1+\rm x^4) \frac{\rm d^2}{\rm {dx}^2} + 2(1-2j)g
\rm x^3 \frac{\rm d}{\rm {dx}} + 2j(2j-1)g \rm x^2.
\end{equation}
The generators $\rm J_{\pm}$ in (3.13) are nothing else but the {\it
second} power of the corresponding {\it linear} $sl(2,\Re)$-generators
(2.13). The Higgs {\it and} $sl(2,\Re)$ algebras are simultaneously
spectrum generating algebras of this application developed in a
$(2j+1)$-dimensional space. This context exactly corresponds to the
Karassiov-Klimov model and its deduced supersymmetric features [18]:
the double degeneracy of the energy eigenvalues is clearly explained
here through the above powers of the linear
$sl(2,\Re)$-generators as well as the possible construction of two
supercharges [30].
\par
Up to the fact that the realization is infinite-dimensional, the
results (3.10) and (3.11) lead once again to the conclusion that we
are dealing with second powers of the generators of the linear
$sl(2,\Re)$-algebra, this context being characterized by the Hamiltonian
\begin{equation}
H = 2 \omega j + g (1+\rm x^4) \frac{\rm d^2}{\rm {dx}^2} + (4j+6)g
\rm x^3 \frac{\rm d}{\rm {dx}} + (4j^2+10j+6)g \rm x^2.
\end{equation}
\par
The third case corresponding to eqs. (3.8) and (3.9) is more
interesting. Indeed, we get the Hamiltonian
\begin{equation}
H = 2 \omega j + g (1+\rm x^4) \frac{\rm d^2}{\rm {dx}^2} + 4g
\rm x^3 \frac{\rm d}{\rm {dx}} - (2j^2+2j-\frac{3}{2}) g \rm x^2.
\end{equation}
Here we notice that our raising operator $\rm J_+$ cannot be related to
the generators of the linear $sl(2,\Re)$ algebra. Such
a family of applications is thus characterized by the (cubic) Higgs
algebra seen as the spectrum generating algebra of interest, this
context being typical once again of infinite-dimensional
realizations. Moreover, let us point out that this Hamiltonian (3.16)
is a Hermitean operator while the previous expressions (3.14) and
(3.15) were not.
\par
The above discussion can evidently be completed by the $N=1$-context
through the values (2.26) with $\beta$ given once again by eq. (3.7).
Here also, we have found (finite- or infinite-dimensional)
realizations entering a (nonhermitean) Hamiltonian admitting the
Higgs algebra as spectrum generating algebra without any connection
with the linear case or its powers. It reads
\begin{eqnarray}
&&H = 2 \omega j + 16g \rm x^4 \frac{\rm d^3}{\rm {dx}^3} + 64g(\rm c
+1) \rm x^3 \frac{\rm d^2}{\rm {dx}^2}\nonumber \\
&&+ g(1+16(6 \rm c^2 +6\rm c -
\frac{1}{2}j^2-\frac{1}{2}j+\frac{9}{4})\rm x^2)\frac{\rm d}{\rm
{dx}}\nonumber \\
&&+ 32g\rm c(2 \rm c^2 -
\frac{1}{2}j^2-\frac{1}{2}j+\frac{1}{4})\rm x
\end{eqnarray}
and we notice that it includes third orders of derivatives as well as
complex energies so that is is not appealing in connection with
physical applications.
\par
In conclusion of this subsection, let us point out that the above
discussion gives us some {\it academic} results ensuring the role of
spectrum generating algebras played by the Higgs algebra in the models
characterized by the Hamiltonians (3.14)-(3.17) but, through their
explicit forms, we immediately see that such Hamiltonians cannot lead
us, in the infinite-dimensional cases, to Schr\" odingerlike equations
easy to handle. For such reasons we want to exploit another
Hamiltonian (3.4) as given in the following subsection.
\subsection{Towards the quadratic algebra}
\hspace{5mm}
Instead of $r=s=2$, let us study the relations (3.1)-(3.5) when $r=1$,
$s=2$, resulting in the Hamiltonian
\begin{equation}
H = (\omega_1 + \omega_2) \rm R_0 + (2\omega_1  - \omega_2 ) \rm J_0 +
g \rm J_+ + \bar g \rm J_-,
\end{equation}
\begin{equation}
\rm J_0 = \frac{1}{3}(N_1 - N_2) \; , \; \rm R_0 = \frac{1}{3} (N_1 +
2 N_2).
\end{equation}
Here the supplementary commutation relation becomes
\begin{equation}
[\rm J_+ , \rm J_-] = \rm P (\rm J_0) = 12 \rm J_0^2 -3\rm R_0 (\rm
R_0 + 1)
\end{equation}
so that we have to consider the polynomial (2.17) for
\begin{equation}
\alpha_1 = -3j (j + 1)\; , \; \alpha_2 = 0 \; , \; \alpha_3 =
12.
\end{equation}
Let us stress that this quadratic algebra is thus a
$W_3^{(2)}$-algebra [3] of special interest appearing, here, in
connection with {\it quantum optical models}.
\par
After calculations, it appears (like for Subsection 2.B) that
the $N=3$-context is impossible but the others $N=2$ and $N=1$ are
possible and lead to three different differential realizations. When
{\it N=2},
\par
if $c = - j$, we get
\begin{equation}
\rm J_0 = \frac{1}{2} (\rm D - j)\; , \; \rm J_- = \frac{\rm
d^2}{\rm {dx}^2} \; , \; \rm J_+ = - \frac{1}{2} \rm x^2 (\rm D -
3j)
\end{equation}
\par
and if $c =  j + 1$, we obtain
\begin{equation}
\rm J_0 = \frac{1}{2} (\rm D + j + 1)\; , \; \rm J_- =
\frac{\rm d^2}{\rm {dx}^2} \; , \; \rm J_+ = - \frac{1}{2} \rm x^2 (\rm
D + 3j + 3).
\end{equation}
In the case {\it N=1}, the parameter $c$ is arbitrary and we get
\begin{equation}
\rm J_0 =  \rm D + \rm c\; , \; \rm J_- =
\frac{\rm d}{\rm {dx}} \; , \; \rm J_+ = - 4 \rm x (\rm
D^2 + (3 \rm c + \frac{1}{2})\rm D + 3 \rm c^2 - \frac{3j^2}{4} -
\frac{3j}{4}).
\end{equation}
\par
Choosing again $\omega_1 = \omega_2 = \omega$ and $g = \bar g$,
the Hamiltonian (3.18) leads to three specific models in
correspondence with the choices (3.22)-(3.24). We respectively get
\begin{equation}
H_1 = \frac{3}{2} \omega j + \rm g \frac{\rm d^2}{\rm {dx}^2} +
\frac{1}{2} (\omega \rm x - \rm g \rm x^3) \frac{\rm d}{\rm {dx}} +
\frac{3}{2} \rm g j \rm x^2,
\end{equation}
\begin{equation}
H_2 = \frac{1}{2} \omega (1+5 j) + \rm g \frac{\rm d^2}{\rm
{dx}^2} +\frac{1}{2} (\omega \rm x - \rm g \rm x^3) \frac{\rm d}{\rm
{dx}} -\frac{1}{2} \rm g (3j+3) \rm x^2
\end{equation}
and
\begin{eqnarray}
&&H_3 =  \omega (\rm c + 2 j) - 4\rm g \rm x^3\frac{\rm
d^2}{\rm {dx}^2} + ( \rm g +\omega \rm x - 4\rm g (3\rm c +
\frac{3}{2})\rm x^2)\frac{\rm d}{\rm {dx}}\nonumber \\
&&-4 \rm g (3 \rm c^2 -\frac{3j^2}{4}-\frac{3j}{4})\rm x^Ê
\end{eqnarray}
None of these Hamiltonians is Hermitean but $H_1$ and $H_2$ are
$PT$-invariant so that they have real spectra [31]. This eliminates
$H_3$ for evident reasons and we have to enlighten the characteristics
of $H_1$ and $H_2$ in order to determine {\it the} Hamiltonian which
has to play the interesting role in quantum optics.
\par
Let us {\it first} establish through conventional quantum mechanical
methods that the stationary Schr\" odinger equation corresponding to
$H_1$ is of the form
\begin{equation}
(-\frac{1}{2} \frac{\rm d^2}{\rm {dx}^2} + V_1(\rm x)) \psi (\rm x) =
E \rm g^{-1} \psi (\rm x)
\end{equation}
with
\begin{equation}
V_1(\rm x) = \frac{1}{32} \rm x^6 - \frac{\omega}{16\rm g} \rm x^4 +
(\frac{\omega^2}{32\rm g^2} - \frac{3j}{4} - \frac{3}{8}) \rm x^2 -
\frac{\omega}{4\rm g} (3j-\frac{1}{2})
\end{equation}
while the one corresponding to $H_2$ is characterized by the potential
\begin{equation}
V_2(\rm x) = \frac{1}{32} \rm x^6 - \frac{\omega}{16\rm g} \rm x^4 +
(\frac{\omega^2}{32\rm g^2} + \frac{3j}{4} + \frac{3}{8}) \rm x^2 -
\frac{\omega}{4\rm g} (5 j+\frac{1}{2})
\end{equation}
We thus recover two particular forms of the famous potential of degree
6, typical of quasi-exactly solvable equations [24, 25, 32] appearing
now in quantum optics. Due to these results we are able to calculate a
certain number of eigenvalues of the spectrum as well as their
eigenfunctions
\par
More precisely, by applying the techniques of [25], [32] to the sextic
oscillators above, we notice that only the potential $V_1 (\rm
x)$ gives a quasi-exactly solvable Schr\" odinger equation leading to
the existence of $[\frac{3j}{2}]+1$ exact solutions (where $[y]$ refers
to the entire part of $y$) and corresponding to real energy eigenvalues
whose the first ones are
\begin{eqnarray}
&&j=0\; : \; E_0 = 0\; ; \; j=\frac{1}{3}\; : \; E_1 = \omega\; ; \;
j=\frac{2}{3}\; :
\; E_2 = \frac{3}{2}\omega \pm \frac{1}{2}(\omega^2 + 8 \rm
g^2)^{\frac{1}{2}}\; ; \nonumber \\
&& j=1\; : \; E_3 = \frac{5}{2}\omega \pm \frac{1}{2}(\omega^2 + 24 \rm
g^2)^{\frac{1}{2}},\; \; \; etc...
\end{eqnarray}
This leads to the conclusion that, with the differential realizations of
the generators of the {\it quadratic} Lie algebra (3.20), the Schr\"
odinger equation (3.28) with (3.29) is the only one
associated with a realistic quantum optical model.
As a {\it second} point, we also would like to comment about the
supersymmetric properties of the Hamiltonian obtained above. For
this purpose, let us consider (3.25) for $j=0$ and, thus, for the
eigenvalue $E_0=0$, it simplifies to
\begin{equation}
H_1 = \frac{\rm d^2}{\rm {dx}^2} - \frac{1}{16} \rm x^6 +
\frac{\omega}{8\rm g} \rm x^4 + (\frac{3}{4}-\frac{\omega^2}{16\rm
g^2}) \rm x^2 -\frac{\omega}{4\rm g}.
\end{equation}
Interestingly, this operator can be expressed in
terms of a superpotential [28] $W_1(\rm x)$
\begin{equation}
H_+ = H_1 = \frac{\rm d^2}{\rm {dx}^2} - W_1^2 - \frac{\rm dW_1}{\rm
{dx}}
\end{equation}
with
\begin{equation}
W_1(\rm x) = - \frac{1}{4} \rm x^3 + \frac{\omega}{4\rm g} \rm x.
\end{equation}
The superpartner $H_-$ of (3.33) is given [28] by
\begin{equation}
H_- = \frac{\rm d^2}{\rm {dx}^2} - W_1^2 + \frac{\rm dW_1}{\rm
{dx}},
\end{equation}
and surprisingly appears to be related to $H_2$:
\begin{equation}
H_- = H_2 \mid _{n=0}.
\end{equation}
It shows that its nonquasi-exactly solvable context is strongly
related to the {\it missing} fundamental energy $E_0=0$ and that it
does not correspond to the expected results of the quantum optical
model. Let us also point out that these superpartners indicate that
(only) for $n=0$, we describe here an {\it exact} supersymmetry [28].
\par
A final {\it third} property that we want to stress is related to
the realization (2.13) of the linear $sl(2,\Re)$: we have also noticed
that the interesting Hamiltonian (3.25) is once again a simple function
of the three generators given in eq. (2.13) but expressed in terms of a
new variable $z=x^2$. We thus conclude that the linear $sl(2,\Re)$ as
well as quadratic algebras we are considering (this latest way being, by
far, more straightforward) are simultaneously spectrum generating
algebras for the model under study. With $\rm c = - \frac{3j}{4}$ and
the realization (2.13) given by
\begin{equation}
\rm j_+ = -z^2 \frac{\rm d}{\rm dz} + \frac{3j}{2} z\; , \; \rm j_- =
\frac{\rm d}{\rm dz}\; , \; j_0 = z\frac{\rm d}{\rm dz} - \frac{3j}{4},
\end{equation}
we find
\begin{equation}
H_1 = \frac{9}{4} \omega j + \omega \rm j_0 + \rm g \rm j_+ + (3j+2)
\rm g \rm j_- + 4 \rm g \rm j_0 \rm j_-.
\end{equation}
\par
The above study shows the real importance of those nonlinear
structures we have put in evidence in connection, here, with quantum
optical models. Let us finally notice that another application of
nonlinear algebras has recently appeared [33] in the context of the
Calogero-Sutherland model [34].   \\

{\bf ACKNOWLEDGEMENTS}
\par
We want to thank the referees for questioning our first results and
for leading us to more interesting new ones.

\end{document}